\documentstyle[12pt,epsf]{article}

\renewcommand{\baselinestretch}{1.1}
\def\beq{\begin{equation}}
\def\eeq{\end{equation}}
\def\beqa{\begin{eqnarray}}
\def\eeqa{\end{eqnarray}}
\def\rd{{\rm d}}
\def\pt{p_\bot}
\def\as{\alpha_s}
\def\eps{\epsilon}
\def\lsim{\raisebox{-3pt}{$\>\stackrel{<}{\scriptstyle\sim}\>$}}
\newcommand{\eqr}[1]{~(\ref{#1})}

\begin{document}

\begin{titlepage}

\begin{flushright}
\begin{tabular}{l}
                FNT/T-93/43 \\
                hep-ph/9311260 \\
                October 1993
\end{tabular}
\end{flushright}
\vspace{2.5cm}

\begin{center}
{\huge Large $p_\bot$ Hadroproduction \\
\vspace{.2cm}
of Heavy Quarks} \\
\vspace{2.5cm}
{\large Matteo Cacciari$^a$ and Mario Greco$^{a,b}$} \\
\vspace{.5cm}
{\sl $^a$Dipartimento di Fisica Nucleare e Teorica\\
Universit\`a di Pavia, Pavia, Italy \\
\vspace{.2cm}
$^b$INFN, Laboratori Nazionali di Frascati, Frascati, Italy }\\
\vspace{2.5cm}

\begin{abstract}
The production of heavy quarks at large $\pt$ ($\pt\gg m$) in hadronic
collisions
is considered. The analysis is carried out in the framework of perturbative
fragmentation functions, thereby allowing a resummation at the NLO level of
final state large mass logarithms of the kind $\log(\pt/m)$. The case of
$b$-quark production is considered in detail. The resulting theoretical
uncertainty from factorization/renormalization scales at large $\pt$ is found
to
be much smaller than that shown by the full $O(\as^3)$ perturbative
calculation.
\end{abstract}

\end{center}
\renewcommand{\baselinestretch}{1}

\vfill
\hrule
\small
\vspace{-.5cm}
\begin{tabbing}
E-mail addresses: \= cacciari@pv.infn.it \\
		  \> greco@lnf.infn.it
\end{tabbing}
\end{titlepage}

\section{Introduction}

Much of the theoretical and experimental interest has been recently devoted to
the study of the production of heavy quarks in hadronic collisions. On the
theoretical side
the calculation in perturbative QCD of the differential and total cross
sections
to order $\as^3$ has been performed by two groups~\cite{NDE1,NDE2,wim1,wim2},
providing a firm basis for a
detailed study of the properties of the bottom and charm quarks, and leading to
reliable predictions for the production rate of the top
quark~\cite{altarelli,laenen}.

	These results do however present a non-negligible residual
renormalization/factorization scale dependence, particularly at large $\pt$.
Furthermore, the validity of this next-to-leading (NLO) $O(\as^3)$ calculation
is limited when $\pt \gg m$,
$m$ being the large quark mass, by the appearance of potentially large
logarithms
of the type $\log(\pt/m)$, which have to be resummed to all orders. The
physical
reason for that is quite clear. For example, terms of order
$(\as^3)\log(\pt/m)$
or $(\as^4)\log(\pt/m)^2$ are simply related to the mass singularities
originating from collinear configurations when $m \to 0$ for fixed $\pt$.
The theoretical uncertainty associated to those corrections
has been roughly estimated in~\cite{NDE2}.
Whereas for top quark production this uncertainty is irrelevant, this is not
the
case for the production of bottom and charm quarks at large $\pt$, leading to
relevant phenomenological consequences.

	In the present paper we consider a solution to this problem, namely we
study the behaviour of the differential cross section for the production of
heavy quarks in the limit $(\pt/m) \gg 1$, to NLO accuracy.
To this aim we follow an
approach based on the properties of fragmentation of a generic parton $p~(p =
q,g,Q)$ in the heavy quark $Q$, after the parton has been produced
inclusively in the
hard collision of the two initial hadrons. The basic formula is represented by
eq.\eqr{eq:fact}, where the partonic cross sections $\hat\sigma_{ij\to kX}$ at
$O(\as^3)$ have been
given in ref.~\cite{aversa} in the massless quark limit.
$\hat\sigma_{ij\to kX}$ introduces an explicit
dependence on $\pt$ and on renormalization/factorization mass scales.
The dependence on the heavy quark mass is
then obtained through the fragmentation function of the parton $p \to Q + X$,
evolved at NLO accuracy from an initial scale $\mu_0 \sim m$ (see below)
to $\mu \sim \pt$.
This approach explicitly resums potentially large terms ot the kind $[\as
\log(\pt/m)]^n$, giving a better description of the theoretical predictions at
large $\pt$. Indeed the corresponding uncertainty is quite reduced in this
region with
respect to the perturbative result, due to a significantly smaller sensitivity
to the relevant scales. On the other hand, because of the massless
limit used for the $O(\as^3)$ kernel cross sections $\hat\sigma_{ij\to kX}$,
this approach
does not allow to recover in a simple way the limit $\pt\lsim m$ of
the perturbative calculation.

In this paper we restrict our approach to $b$-quark production,
the application to
charm production will be given elsewhere. In Sect.~\ref{sect2} we give the
general formalism. In Sect.~\ref{sect3} we first discuss the basic
ingredients necessary to obtain our
NLO cross sections and then we present our results. We compare with the
perturbative calculation of ref.~\cite{NDE2} and study in detail the
dependence on the various
mass scales entering the problem. Our conclusions are finally given in
Sect.~\ref{sect4}.

\section{The fragmentation function approach}
\label{sect2}

According to factorization theorems~\cite{css} the cross section for the
inclusive hadroproduction of a hadron at high transverse momentum, i.e.
for the process
\begin{eqnarray*}
H_1 + H_2 \to H_3 + X \nonumber
\end{eqnarray*}
can be written as
\beq
\rd\sigma = \sum_{i,j,k}\int \rd x_1\, \rd x_2\,
\rd x_3\, F_{H_1}^i(x_1,\mu_F) F_{H_2}^j(x_2,\mu_F)
\rd\hat\sigma_{ij\to kX}(x_1,x_2,\mu_R,\mu_F) D_k^{H_3}(x_3,\mu_F)
\label{eq:fact}
\eeq
As usual, the $F$'s are the distribution functions of the partons in
the colliding
hadrons, $\hat\sigma$ is the kernel cross section and $D$ is the fragmentation
function of the observed hadron. The factorization mass scales $\mu_F$ of
the structure
and fragmentation functions are assumed to be equal for the sake of simplicity.
$\mu_R$ is the renormalization scale.

Due to the presence of collinear singularities both in the initial and
final state this process is not fully predictable by QCD itself. We can
actually
calculate the kernel cross section
and the evolution of the structure
and fragmentation functions, but we have to rely on some
phenomenological input to obtain the latter at some
given initial scale.

This situation changes drastically when we come to consider the inclusive
production of a heavy quark. In this case its mass, being finite
and considerably greater than $\Lambda$, makes the perturbative expansion
feasible and prevents collinear singularities from appearing in the splitting
vertices which involve the heavy quark.

Having this in mind two approaches can be pursued in the calculation of heavy
quark production.

The first one is to directly calculate in perturbation theory the
process $\rd\hat\sigma_{ij\to QX}$, Q being the heavy quark and
$i,j$ the initial state {\it light} partons (i.e. light quarks and gluons).
This kernel cross section will then be convoluted with initial state structure
functions only, the final state showing no singularities of any kind. This
approach has been followed in the past~\cite{NDE1,NDE2,wim1,wim2},
providing a full perturbative $O(\as^3$) calculation. In the
following we shall use for comparisons the results of Nason, Dawson and Ellis,
and refer to them as NDE. In this perturbative approach, as stated earlier,
terms of the kind $\as\log(\pt/m)$ will appear. They are remnant of
the collinear singularity screened by the finite quark mass. As noted
in ref.~\cite{NDE2} they can
grow quite large at high tranverse momenta, thereby spoiling the validity of
the expansion in $\as$. Therefore they have to be summed to all orders.

The alternative way is to consider that when a quark, of whichever
flavour, is produced at very high transverse momentum $\pt \gg m$ its mass
plays almost no role at all in the scattering process. This is to say that mass
effects in
the kernel cross section are suppressed as power ratios of mass over the scale
of the process. We can therefore devise a picture in which all quarks are
produced in a massless fashion at the high scale $\mu_F \sim p_\bot \gg m$
and only successively, as their virtuality
decreases, they can {\em fragment} into a massive heavy quark. The cross
section
can therefore be described by a formula analogous to eq.\eqr{eq:fact},
with $H_3 = Q$.
The key difference
to the hadron production case considered in eq.\eqr{eq:fact} is that initial
state conditions for the heavy quark fragmentation functions are now calculable
from first principles in QCD (hence the definition of ``perturbative''
fragmentation functions, PFF) and do not have to be taken from experiment.

Actually, the following set of next-to-leading initial state conditions can be
obtained~\cite{melenason} in the $\overline{MS}$ scheme for the fragmentation
function of a heavy quark, gluon and light quark respectively, in the heavy
quark $Q$
\beqa
&&D_Q^Q(x,\mu_0) = \delta(1-x) + {{\as(\mu_0) C_F}\over{2\pi}}\left[
{{1+x^2}\over{1-x}}\left(\log{{\mu_0^2}\over{m^2}} -2\log(1-x)
-1\right)\right]_+ \label{DQQ} \\
&&D_g^Q(x,\mu_0) = {{\as(\mu_0) T_f}\over{2\pi}}(x^2 + (1-x)^2)
\log{{\mu_0^2}\over{m^2}} \label{DgQ} \\
&&D_{q,\bar q,\bar Q}^Q(x,\mu_0) = 0 \label{DqQ}
\eeqa
where $\mu_0$ must be taken of the order of the heavy quark mass.

Then using the usual Altarelli-Parisi evolution equations at NLO accuracy one
finds the fragmentation functions set at any desired factorization scale
$\mu_F$. An important feature of this approach can now be appreciated.
The ``almost-singular'' logarithmic term $\log\left( {\pt/m} \right)$
splits into two, as follows.

A $\log\left( {\pt/\mu_F} \right)$ will be found in the
kernel cross section $\hat\sigma$ which has no dependence on the
heavy quark mass, according to the assumption that it is
produced in a {\em massless} way. Moreover, by choosing $\mu_F \sim p_\bot$ it
will not contain large logarithms and its perturbative expansion will behave
correctly.

The remaining part of the log will instead be lurked into the
fragmentation function $D(x_3,\mu_F)$. The large $\log(\mu_F/\mu_0)$ is
resummed to
all orders by the evolution equations, and only the small $\log(\mu_0/m)$
provided by the initial state condition is treated at fixed order in
perturbation theory. Therefore one expects a better control of the theoretical
uncertainty at large $\pt$. On the other hand, for $\pt \lsim m$ the
fragmentation
approach does not allow to recover easily the perturbative result, which, of
course, holds exactly.

\section{NLO Cross Section Evaluation}
\label{sect3}

In order to implement the ``perturbative fragmentation function (PFF)
approach'' at a numerical level we need
four ingredients, which are all available at the next-to-leading level:
\begin{itemize}
\item[{\em i})] the distribution functions of any parton (including the heavy
flavour in question)
in the hadrons (proton or antiproton), evolved at NLO accuracy. All modern sets
satisfy this requirement. An important point must however be made clear. A
heavy quark present in the initial state can be directly brought to the final
state where it is fragmented to
the detected heavy flavour through the $D_Q^Q$, and therefore with high
probability (see below). This means that the resulting cross section
is particularly
sensitive to the overall heavy flavour content of the colliding hadrons. In the
Parton Distribution Functions (PDF) sets available this content
is generated through perturbative gluon
splitting above a given threshold. The total yield will therefore depend on the
choice made. For instance, the HMRS set~\cite{HMRS} takes $F_b(x,2m_b) = 0$
as initial
condition, whereas the MT~\cite{MT} and the CTEQ~\cite{CTEQ} ones choose,
according to ref.~\cite{ct},
$F_b(x,m_b) = 0$. The resulting different bottom distributions are plotted in
figure~\ref{bpdf}: the MT and CTEQ bottom contents are consistently higher
than the HMRS
one, the discrepancy growing larger as the HMRS threshold value is approached.

\item[{\em ii})] the kernel cross section for the scattering of any two
massless partons into another massless parton. This calculation
is provided at the NLO
in various renormalization/factorization schemes in ref.~\cite{aversa}. The
massless limit approximation used in these $O(\as^3)$ kernel cross sections
will not allow to evaluate the heavy quark cross section for $\pt\lsim m$.

\item[{\em iii})] the next-to-leading expression for the strong running
coupling:
\begin{eqnarray*}
& & \as (\mu^2) = \frac{1}{b \ln (\mu^2/\Lambda^2)} \; \left[
1 - \frac{b'}{b} \: \frac{\ln \ln (\mu^2/\Lambda^2)}{\ln (\mu^2
/\Lambda^2)} \right] \nonumber\\ \vbox{\vspace{.4cm}}
& & b = \frac{33-2N_f}{12 \pi}, \; \; b'= \frac{153 -19 N_f}{24 \pi^2}\nonumber
\end{eqnarray*}
The QCD scale $\Lambda$ is set at the value fixed by the structure function set
used and the number of active flavours $N_f$ is set at 5 above the appropriate
threshold.

\item[{\em iv})] the fragmentation functions of any parton into the heavy
flavour. They are obtained by evolving the initial conditions given above
(eqs.~\ref{DQQ},\ref{DgQ},\ref{DqQ}) with NLO
accuracy~\cite{cfp}. This is done through numerical inversion of the
Mellin moments of
the evolved distributions\footnote{We thank Paolo Nason for having provided us
with the FORTRAN code for this.}, and a typical result for
bottom fragmentation is
depicted in figure~\ref{bfrag}, together with the uncertainty given by the
choice of the initial condition scale $\mu_0$ around $m_b$. The implication
of these ideas to LEP hadronic data is discussed in ref.~\cite{colangelo}.
\end{itemize}

With these four ingredients at our disposal we can now
evaluate the cross section for the high $\pt$ inclusive hadroproduction of a
bottom quark. This will be done by performing numerically the triple
integration
of eq.\eqr{eq:fact}. It must be noted that this integration is not as
straightforward as the one usually performed when calculating hadron
distributions. Indeed while the hadronic fragmentation functions fall smoothly
to zero as $x$ approaches one, now the $D_Q^Q$ is instead singular in the
$x\to 1$ limit, due to unresummed Sudakov terms.
This makes the most usual integration routines to fail in calculating both the
integral and the associated numerical error, the latter being heavily
underestimated.

More in detail this problem has been overcome by suitably rearranging the
integration in such a
way that the numerical integration could be reliably performed. By rewriting
eq.\eqr{eq:fact} in a more compact form, and dropping all unnecessary symbols,
we have
\beqa
\sigma &=& \int^1_{x_3^{min}}\rd x_3\int\rd x_1 \rd x_2 \sum_{ijk} F_i F_j
\hat\sigma_{ijk}(x_3) D_k(x_3) = \nonumber \\
&=& \int^1_{x_3^{min}}\rd x_3\int\rd x_1 \rd x_2 \sum_{\stackrel{
\scriptstyle{ijk}}{k\neq Q}}
F_i F_j \hat\sigma_{ijk}(x_3) D_k(x_3) + \nonumber \\
&+& \int^{1-\eps}_{x_3^{min}}\rd x_3\int\rd x_1 \rd x_2 \sum_{ij} F_i F_j
\hat\sigma_{ijQ}(x_3) D_Q(x_3) + \nonumber \\
&+& \int_{1-\eps}^1\rd x_3 \left[{{\int\rd x_1 \rd x_2 \sum_{ij} F_i F_j
\hat\sigma_{ijQ}(x_3)}\over{x_3}} -  {{\int\rd x_1 \rd x_2 \sum_{ij} F_i F_j
\hat\sigma_{ijQ}(1)}\over{1}}\right] x_3 D_Q(x_3) + \nonumber \\
&+& \int\rd x_1 \rd x_2 \sum_{ij} F_i F_j
\hat\sigma_{ijQ}(1)\int_{1-\eps}^1\rd x_3~x_3D_Q(x_3)
\eeqa
The last integral over $x_3$ is  then evaluated by resorting to the Mellin
moments\footnote{We
recall that the Mellin moments are defined by
\begin{eqnarray*}
{\widetilde D}(N) = \int_0^1 \rd x~x^{N-1} D(x)\nonumber
\end{eqnarray*}
}
of the evolved fragmentation function:
\beq
\int_{1-\eps}^1\rd x_3~x_3D_Q(x_3) = {\widetilde D}_Q(2) -
\int_0^{1-\eps}\rd x_3~x_3D_Q(x_3)
\label{coeff}
\eeq

The result should of course be independent of the choice of $\eps$, and we have
verifyed that this is indeed the case for $0.9\lsim\eps\lsim 0.99$.

Figure~\ref{1800} shows a comparison of the perturbative result of
ref.~\cite{NDE2} with our calculations for two different PDF sets, at
the Fermilab energy of 1800 GeV, for $\mu = \sqrt{m_b^2 + \pt^2}$. We can see
that in the high $\pt$ region the perturbative cross section is quite sensitive
to the structure function set choice, the MT one giving a markedly lower
result.
The opposite happens in the PFF approach, which becomes very
sensitive in the
low $\pt$ region which, as stated above, is plagued by the neglect of
the heavy quark mass
terms in the partonic cross sections. This pattern is due to the fact
that the NDE calculation only uses the light quark content from the
structure functions, since the heavy one is produced at the purely perturbative
level. This introduces a mismatch (albeit of higher order) as the
evolution had been performed by taking
into account all flavours, which get mixed through gluon exchange.
Our calculation, on the other
hand, treats all quarks on the same ground and uses the full information of the
PDF set. This leads to
substantially identical predictions obtained in the high $\pt$ region with the
two PDF sets. On the contrary, at low $\pt$ the different threashold behaviour
of the two sets in the heavy quark content is responsible for the discrepancy
between the two results.

Figure~\ref{630}a,b show the same kind of comparison at 630 GeV and 16 TeV.
The behaviour
is practically identical, once a rescaling on the $\pt$
axis has been considered to account for the different cm energies.

One further comparison can also be made between the MT-B2 and the CTEQ1M sets,
which have a similar threshold behaviour for the heavy quark flavour.
As can be seen in figure~\ref{1800cteq}a,b the theoretical uncertainty due to
the structure functions set choice is once again smaller in the fragmentation
function approach.

Next we consider the dependence on the choice of
the renormalization/factorization mass scale $\mu$.
Figure~\ref{1800sc_mt}a(b) shows, at 1800 GeV and with the MT-B2 (HMRS-B)
set, the theoretical uncertainty resulting from the variation of the
factorization and renormalization mass scales
between $\mu_{ref}/2$ and $2\mu_{ref}$, where
$\mu_{ref}$ is defined as $\sqrt{m_b^2 + \pt^2}$ and we have taken $\mu = \mu_F
= \mu_R$. As expected the band of the perturbative NDE
calculation is sensibly larger than ours, showing the improvement brought by
the
resummation of the large logarithms of $\pt/m_b$. The drop at low $\pt$
in the prediction of the HMRS-B set with $\mu = \mu_{ref}/2$, which
considerably  broadens our band in that region, is due to the very poor heavy
quark content of this set at such a low scale (see Fig.~\ref{bpdf}).
All these features can be better appreciated in
figure~\ref{scale}, where the cross section at 1800 GeV with the MT-B2 set
is plotted, at fixed $y$ and $\pt$, as a
function of $\mu = \xi\mu_{ref}$, for $\xi$ varying between 0.25 and 4.
This figure also shows a comparison with
the factorized calculation with a Born (i.e. LO) cross section kernel
(but with
two-loop $\as$ and NLO structure and fragmentation functions).
As expected, the lack of the next-to-leading terms strongly enhances
the scale dependence. The similarity between the NDE result and the
fragmentation function approach with Born kernel cross section is striking. The
small scale sensitivity of our full NLO calculation shows that the
fragmentation/renormalization scale dependence is a real $O(\as^4)$ effect,
whereas in the perturbative approach (e.g. NDE) the presence of large
$\log(\pt/m)$ results in an effective $O(\as^3)$ dependence.

Once again, the results at 630 GeV greately mimick those at 1800 GeV (see
figure~\ref{630sc_mt}).

One further theoretical uncertainty must be considered, namely the one related
to the freedom in the choice of the initial condition scale $\mu_0$ in
the fragmentation
functions evolution. Figure~\ref{1800mu0} shows that the cross section is
marginally affected by a change of $\mu_0$ in the range $m_b/2 \to 2m_b$,
particularly in the high $\pt$ region. This leads to the conclusion
that the PFF approach has still a smaller overall
theoretical uncertainty
also after this qualitatively new feature has been taken into account.

Finally, we show in figs.~\ref{channels}a,b the relative importance of the
various partonic subprocesses contributing to the cross section for two sets of
PDFs. In the case of the MT-B2 set (similar results are obtained with the
CTEQ1M
one) the $Qg$ scattering clearly dominates al low $\pt$, whereas at large $\pt$
it becomes comparable to other channels. Comparing with the HMRS-B set we see
that the different threshold behaviour of the heavy quark structure function
in this latter set, repetedly pointed
out before, is responsible for the lower contribution of the $Qg$ and $Qq$
channels.

We can now turn to quantities of more direct experimental interest and see
how the advantages of our approach are reflected onto them. Namely,
we will consider the total cross section for one particle inclusive heavy quark
production, integrated above a given $\pt^{min}$ and within a rapidity region
$|y| < y_{max}$. Only the variation of the theoretical prediction due to
changes in the
factorization/renormalization scale and in the PDF set used will be studied.
Other possible sources of uncertainties, aside the change of $\mu_0$ which has
been shown to be almost neglibible, are the value of the QCD scale
$\Lambda$ and of the bottom mass $m_b$. They should however be common to both
the perturbative and the fragmentation function approach, and have been
studied in detail in ref.~\cite{NDE2}.

Figures~\ref{1800_hmrs_tot}, \ref{1800_mt_tot} and \ref{1800_cteq_tot}
show the total cross section
$\sigma(\pt > \pt^{min}, |y| < 1)$ at a cm energy of 1800 GeV. The HMRS-B,
MT-B2
and CTEQ1M sets have been used respectively, and the curves obtained with $\mu
= 2\mu_{ref}$ and $\mu = \mu_{ref}/2$ are plotted. We can see that the features
of the differential cross sections previously discussed translate almost
directly into the total cross section. In particular, the scale dependence of
our result is once again remarkably smaller than the one of the
perturbative calculation, especially in the high $\pt$ region. The broadening
of
the band in the low $\pt$ region than can be observed when the HMRS-B or the
CTEQ1M set is used with the scale choice $\mu = \mu_{ref}/2$ is once again due
to the different heavy quark content at low scales (see fig.~\ref{bpdf}).

The overall smaller theoretical uncertainty of the PFF
result can finally be
appreciated in fig.~\ref{final}, where the highest and lowest predictions of
the
two approaches, out of the six curves previously considered, have been
plotted. Note that we have not considered in detail the uncertainties for
$\pt\lsim 10$~GeV. The CDF experimental data~\cite{abe} are also shown.

The same kind of result can also be produced at 630 GeV. Figure~\ref{final630}
compares the band of NDE to our one and to experimental
data by UA1~\cite{albajar}. Also in this case we find a sizeable reduction
in the uncertainty of the theoretical prediction.

\section{Conclusions}
\label{sect4}

In this paper we have discussed the production of heavy quarks at large $\pt$
($\pt \gg m$) in hadronic collisions, in order to make the theoretical
predictions for the differential distributions more reliable. We have studied
this problem in the framework of the perturbative fragmentation functions,
which allow a NLO
evaluation of the potentially large logarithms of the kind $\log(\pt/m)$, which
are resummed to all orders.

Our analysis for the $b$-quark leads to much more stable results with respect
to
changes of the factorization/renormalization scales compared to what is
obtained
in the $O(\as^3)$ perturbative calculation. Also the theoretical uncertainty
related to different choices of PDF sets is reduced. Other possible sources of
uncertainties, like the scale $\mu_0$ of the initial state condition in the
fragmentation functions evolutions, are negligible. A detailed discussion of
the
differential and integrated cross sections of experimental interest has also
been presented.

\vspace{1cm}
\noindent
We are grateful to P.~Chiappetta, J.-Ph.~Guillet and P.~Nason for their
contribution at the
early stage of this analysis. Useful discussions with M.L.~Mangano are also
acknowledged.


\newpage

\listoffigures
\newpage

\renewcommand{\baselinestretch}{1.}

\newlength{\picsize}
\setlength{\picsize}{8.0 cm}
\newlength{\capsize}
\setlength{\capsize}{7.0 cm}

\let\small\relax
\begin{figure}[th]
\begin{minipage}[t]{\picsize}
\begin{center} \mbox{
\epsfxsize=\picsize
\epsffile{b_pdf.eps}
}
\end{center}
\vspace{-1.5truecm}
\begin{center}
\parbox{\capsize}{
\caption{\label{bpdf}
\small Bottom distributions in hadrons according to HMRS-B (solid
line), MT-B2 (dashed) and CTEQ1M (dotted) sets, at different values of the
factorization scale.}}
\end{center}
\vspace{-.5truecm}
\end{minipage}
\begin{minipage}[t]{\picsize}
\begin{center} \mbox{
\epsfxsize=\picsize
\epsffile{b_frag.eps}
}
\end{center}
\vspace{-1.5truecm}
\begin{center}
\parbox{\capsize}{
\caption{\label{bfrag}
\small Bottom fragmentation functions at $Q = 91$ GeV.}}
\end{center}
\vspace{-.5truecm}
\end{minipage}
\end{figure}
\begin{figure}[th]
\begin{center}
\mbox{
\epsfxsize=\picsize
\epsffile{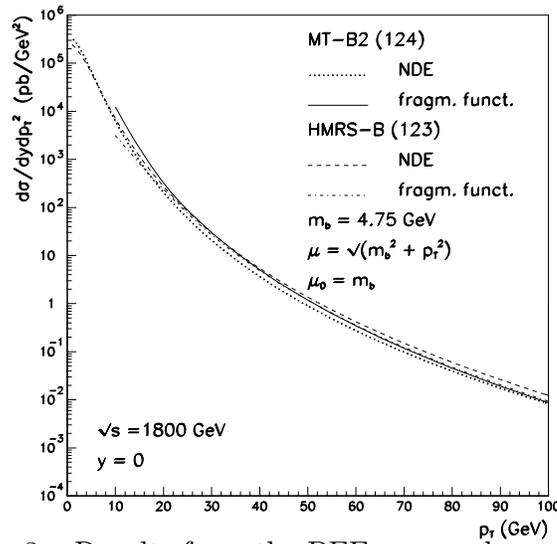}
}
\end{center}
\vspace{-1.5truecm}
\begin{center}
\parbox{10cm}{
\caption{\label{1800}
\small Results from the PFF approach compared to the
perturbative prediction of NDE at 1800 GeV.}}
\end{center}
\vspace{-.5truecm}
\end{figure}
\begin{figure}[th]
\begin{center}
\begin{minipage}[t]{\picsize}
\epsfxsize=\picsize
\epsffile{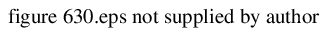}
\end{minipage}
\begin{minipage}[t]{\picsize}
\epsfxsize=\picsize
\epsffile{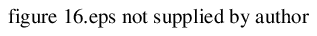}
\end{minipage}
\end{center}
\vspace{-1.5truecm}
\begin{center}
\parbox{10cm}{
\caption{\label{630}
\small Results from the PFF approach compared to the
perturbative prediction of NDE, with the MT-B2 and the HMRS-B sets,
at 630 GeV and 16 TeV.}}
\end{center}
\vspace{-.5truecm}
\end{figure}
\begin{figure}[th]
\begin{center} \mbox{
\epsfxsize=\picsize
\epsffile{1800cteq.eps}
\epsfxsize=\picsize
\epsffile{16cteq.eps}
}
\end{center}
\vspace{-1.5truecm}
\begin{center}
\parbox{10cm}{
\caption{\label{1800cteq}
\small Results from the PFF approach compared to the
perturbative prediction of NDE, with the MT-B2 and CTEQ1M sets, at 1800 GeV and
16 TeV.}}
\end{center}
\vspace{-.5truecm}
\end{figure}
\begin{figure}[th]
\begin{center} \mbox{
\epsfxsize=\picsize
\epsffile{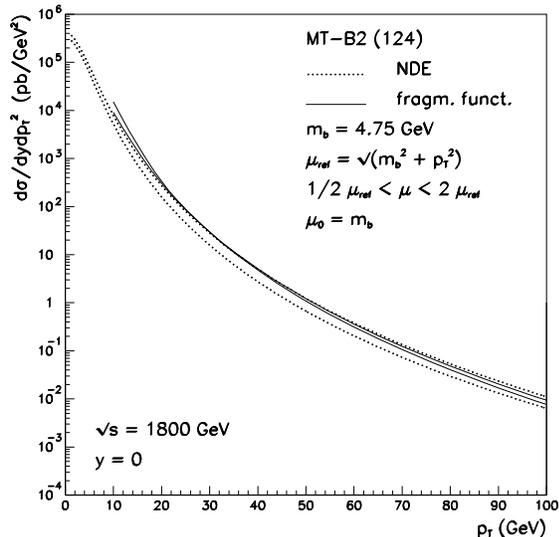}
\epsfxsize=\picsize
\epsffile{1800sc_hmrs.eps}
}
\end{center}
\vspace{-1.5truecm}
\begin{center}
\parbox{10cm}{
\caption{\label{1800sc_mt}
\small Scale dependence at 1800 GeV, with MT-B2 and HMRS-B
structure functions sets.}}
\end{center}
\vspace{-.5truecm}
\end{figure}
\begin{figure}[th]
\begin{minipage}[t]{\picsize}
\begin{center} \mbox{
\epsfxsize=\picsize
\epsffile{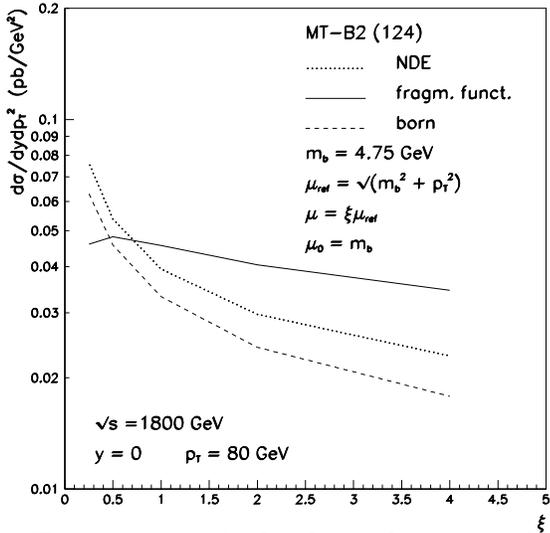}
}
\end{center}
\vspace{-1.5truecm}
\begin{center}
\parbox{\capsize}{
\caption{\label{scale}
\small Scale dependence of the cross section as a function of $\mu =
\xi\mu_{ref}$, at $y=0$ and $\pt$ = 80 GeV. ``NDE'' and ``fragm. funct.'' refer
to the full NLO calculations, whereas ``born'' is the result of the
fragmentation function approach with LO kernel cross section.}}
\end{center}
\vspace{-.5truecm}
\end{minipage}
\begin{minipage}[t]{\picsize}
\begin{center} \mbox{
\epsfxsize=\picsize
\epsffile{630sc_mt.eps}}
\end{center}
\vspace{-1.5truecm}
\begin{center}
\parbox{\capsize}{
\caption{\label{630sc_mt}
\small Scale dependence at 630 GeV.}}
\end{center}
\vspace{-.5truecm}
\end{minipage}
\end{figure}
\begin{figure}[th]
\begin{center} \mbox{
\epsfxsize=\picsize
\epsffile{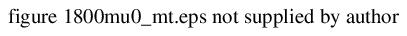}
\epsfxsize=\picsize
\epsffile{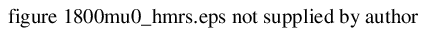}
}
\end{center}
\vspace{-1.5truecm}
\begin{center}
\parbox{10cm}{
\caption{\label{1800mu0}
\small Spread in the theoretical prediction by varying the initial
condition scale $\mu_0$ in the fragmentation function evolution, both with
the MT-B2 and the HMRS-B set.}}
\end{center}
\vspace{-.5truecm}
\end{figure}
\begin{figure}[th]
\begin{center} \mbox{
\epsfxsize=\picsize
\epsffile{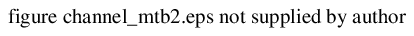}
\epsfxsize=\picsize
\epsffile{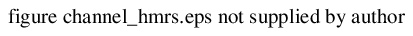}
}
\end{center}
\vspace{-1.5truecm}
\begin{center}
\parbox{10cm}{
\caption{\label{channels}
\small Relative contribution to the cross section of the various partonic
channels, with the MT-B2 and the HMRS-B sets.}}
\end{center}
\vspace{-.5truecm}
\end{figure}
\begin{figure}[th]
\begin{minipage}[t]{\picsize}
\begin{center} \mbox{
\epsfxsize=\picsize
\epsffile{1800_hmrs_tot.eps}
}
\end{center}
\vspace{-1.5truecm}
\begin{center}
\parbox{\capsize}{
\caption{\label{1800_hmrs_tot}
\small Scale dependence of the total cross section above a given $\pt^{min}$
and
within the rapidity region $|y| < 1$ at 1800 GeV, with the HMRS-B set.}}
\end{center}
\vspace{-.5truecm}
\end{minipage}
\begin{minipage}[t]{\picsize}
\begin{center} \mbox{
\epsfxsize=\picsize
\epsffile{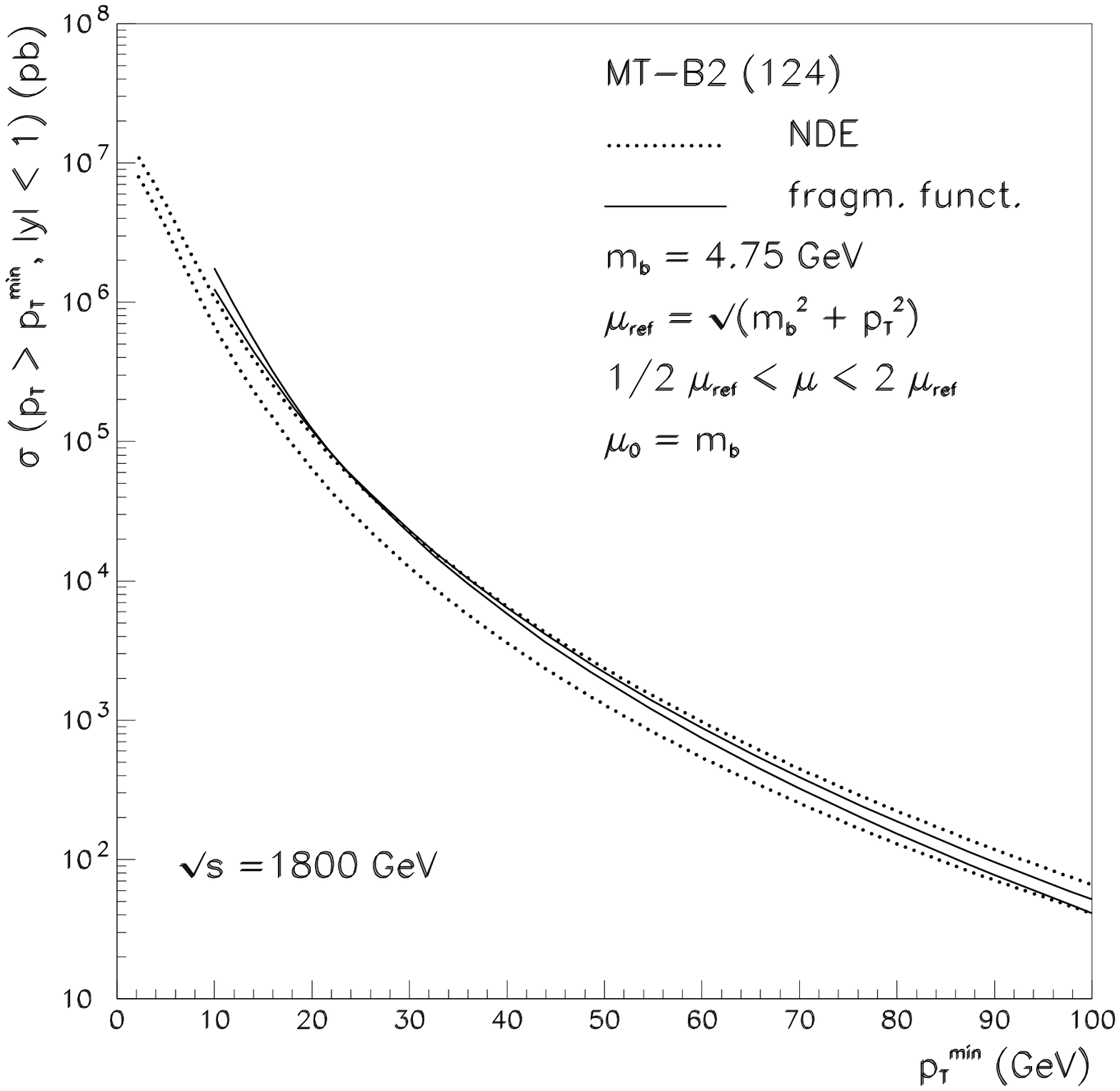}
}
\end{center}
\vspace{-1.5truecm}
\begin{center}
\parbox{\capsize}{
\caption{\label{1800_mt_tot}
\small Scale dependence of the total cross section above a given $\pt^{min}$
and
within the rapidity region $|y| < 1$ at 1800 GeV, with the MT-B2 set.}}
\end{center}
\vspace{-.5truecm}
\end{minipage}
\end{figure}
\begin{figure}[t]
\begin{minipage}[t]{\picsize}
\begin{center} \mbox{
\epsfxsize=\picsize
\epsffile{1800_cteq_tot.eps}
}
\end{center}
\vspace{-1.5truecm}
\begin{center}
\parbox{\capsize}{
\caption{\label{1800_cteq_tot}
\small Scale dependence of the total cross section above a given $\pt^{min}$
and
within the rapidity region $|y| < 1$ at 1800 GeV, with the CTEQ1M set.}}
\end{center}
\vspace{-.5truecm}
\end{minipage}
\begin{minipage}[t]{\picsize}
\begin{center} \mbox{
\epsfxsize=\picsize
\epsffile{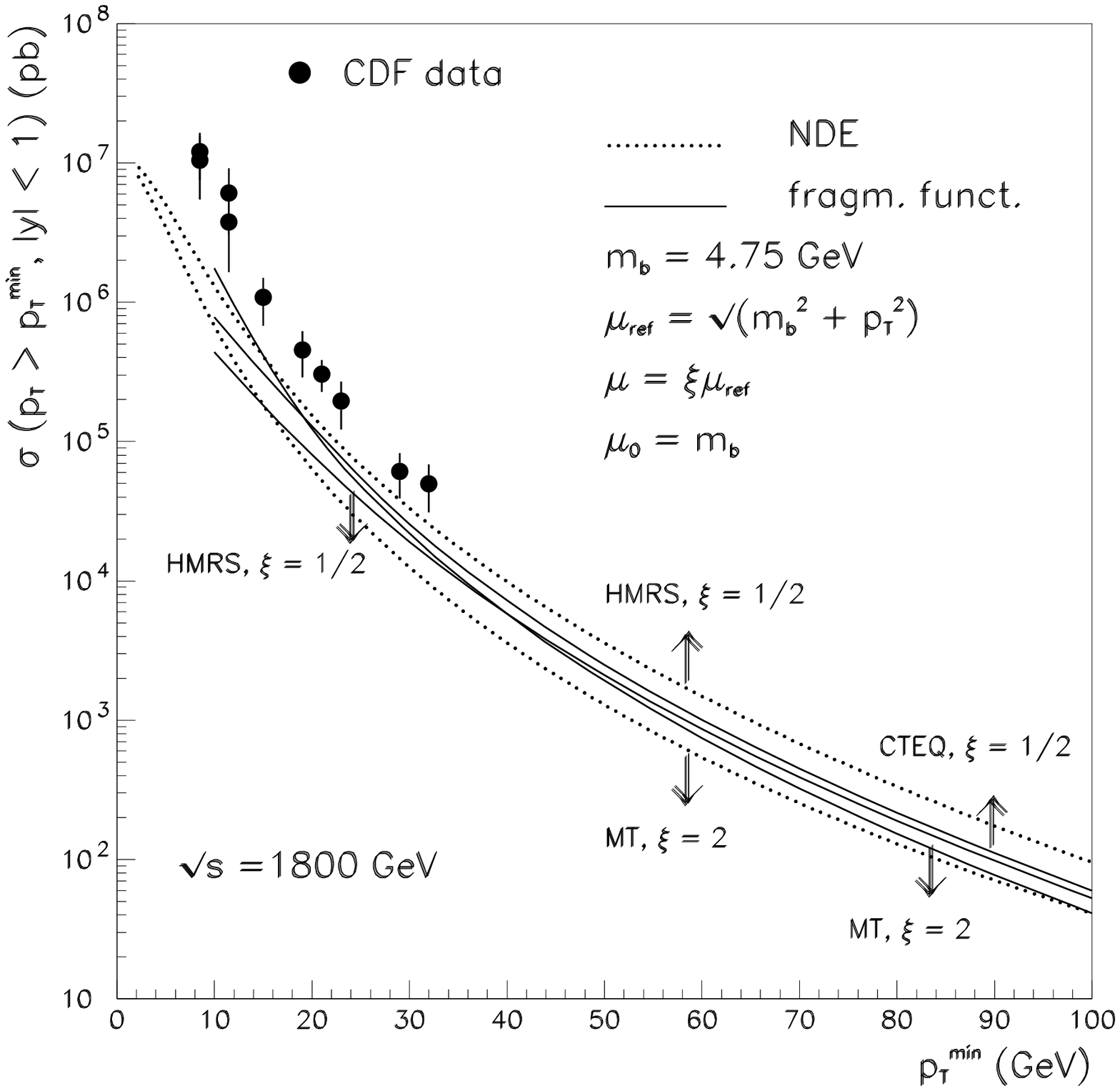}
}
\end{center}
\vspace{-1.5truecm}
\begin{center}
\parbox{\capsize}{
\caption{\label{final}
\small Comparison of the overall theoretical uncertainty of inclusive high
$\pt$ heavy quark production at 1800 GeV in the PFF and
perturbative approaches. CDF experimental data are also shown.}}
\end{center}
\vspace{-.5truecm}
\end{minipage}
\begin{minipage}[t]{\picsize}
\begin{center} \mbox{
\epsfxsize=\picsize
\epsffile{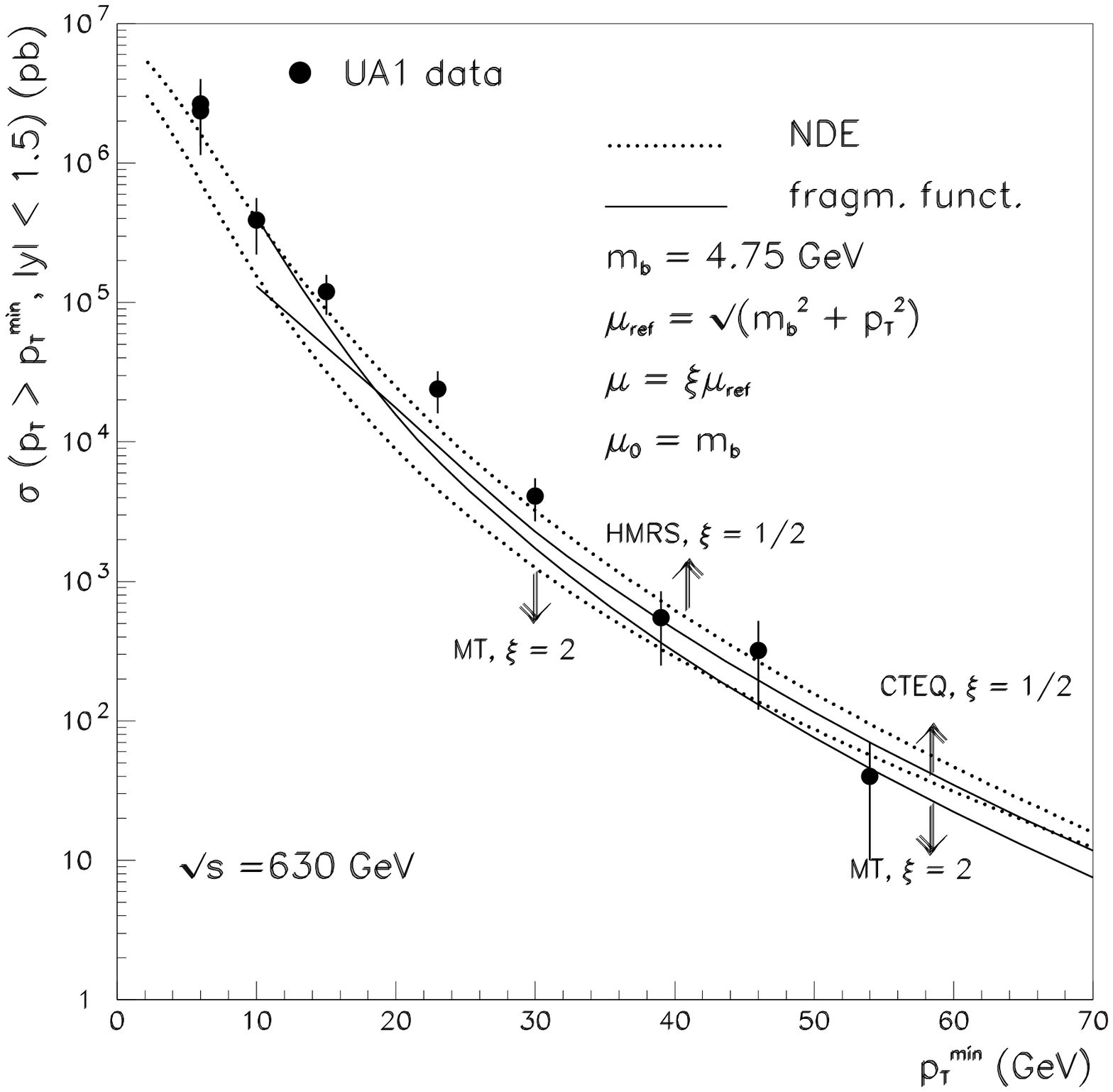}
}
\end{center}
\vspace{-1.5truecm}
\begin{center}
\parbox{\capsize}{
\caption{\label{final630}
\small Comparison of the overall theoretical uncertainty of inclusive high
$\pt$ heavy quark production at 630 GeV in the PFF and
perturbative approaches. UA1 experimental data are also shown.}}
\end{center}
\vspace{-.5truecm}
\end{minipage}
\end{figure}

\end{document}